\documentclass[11pt, a4paper]{article}
\usepackage{jcappub}

\usepackage{color}
\usepackage{booktabs}
\usepackage{multirow}
\usepackage{amssymb}
\usepackage[export]{adjustbox}
\usepackage{floatrow}
\newfloatcommand{capbtabbox}{table}[][3.5in]
\newfloatcommand{capbfigbox}{figure}[][2.3in]
\usepackage{blindtext}
\usepackage{amsmath}
\usepackage{booktabs}
\usepackage{subcaption}
\usepackage{aas_macros}
\usepackage{cleveref}
\usepackage{graphicx}
\usepackage{caption}
\usepackage{subcaption}
\usepackage{orcidlink}

\newcommand{\cm}{\,\text{cm}}

\def\lsim{\mathrel{\raise.3ex\hbox{$<$\kern-.75em\lower1ex\hbox{$\sim$}}}}
\def\gsim{\mathrel{\raise.3ex\hbox{$>$\kern-.75em\lower1ex\hbox{$\sim$}}}}

\begin{document}

\title{Hadronic Emissions from the Microquasar V4641 Sgr, SS433, and its implications in the Diffuse Galactic Emission}

\author[a]{Basanti Paul,\orcidlink{0009-0007-0343-4532}}
\author[b,c]{Abhijit Roy,\orcidlink{0009-0005-2824-655X}}
\author[d, e]{Jagdish C. Joshi,\orcidlink{0000-0003-3383-1591}}
\author[f]{Debanjan Bose \orcidlink{0000-0003-1071-5854}}

\affiliation[a]{School of Astrophysics, Presidency University, Kolkata, India, Pin - 700073}
\affiliation[b]{Gran Sasso Science Institute, via F. Crispi 7 -- 67100, L'Aquila, Italy} 
\affiliation[c]{INFN/Laboratori Nazionali del Gran Sasso, via G. Acitelli 22, Assergi (AQ), Italy}
\affiliation[d]{Aryabhatta Research Institute of Observational Sciences (ARIES), Manora Peak, Nainital 263001, India}
\affiliation[e]{Centre for Astro-Particle Physics (CAPP) and Department of Physics, University of Johannesburg, PO Box 524, Auckland Park 2006, South Africa}
\affiliation[f]{Department of Physics, Central University of Kashmir, Ganderbal, India, Pin -191131}

\emailAdd{basantipaul025@gmail.com, aviatphysics@gmail.com, jagdish@aries.res.in, debaice@gmail.com}

\abstract{
Microquasars (MQs) are Galactic binary systems, consisting of a star and a compact object, a neutron star or a stellar mass black hole, which accretes matter from its companion star and gives rise to relativistic jets. Recent detection of very-high-energy (VHE; $E \gtrsim 100 \,\text{GeV}$) and ultra-high-energy (UHE; $E \gtrsim 100 \,\text{TeV}$) gamma-rays by LHAASO, HAWC and HESS from the MQ V4641 Sgr and SS 433  suggests them as Galactic PeVatrons. In this work, we studied a hadronic origin of the observed TeV-PeV gamma-ray emission from these MQs. We considered the hadronic scenario where the gamma-rays are produced by the interaction of relativistic protons in the MQ jet with the stellar wind. We fitted our model with observed data and constrained physical parameters like the hadronic jet power fraction, the proton spectral index, the maximum proton energy and the jet bulk Lorentz factor. Our best-fit model shows hard proton spectra ($1.84-2.44$) and maximum proton energies between 1 and 5 PeV. We also estimated the all-flavor neutrino fluxes corresponding to the gamma-ray fluxes from the hadronic model and found that V4641 sgr can be detected by next-generation neutrino telescopes like KM3NeT-ARCA and TRIDENT. Furthermore, we modeled a synthetic population of Galactic MQs and estimated their contribution to the diffuse TeV–PeV gamma-ray flux. For the inner Galaxy PSR contribution dominates in the range 10-100 TeV, and above 100 TeV diffused cosmic ray interactions with the molecular clouds is most dominant. We find that a population $\sim 14$ MQs is required to explain the LHAASO data above 100 TeV. For the outer Galaxy, we show that MQs are the dominant class of sources, and we constrain their population $\sim$14. Our findings strongly suggest that MQs are efficient particle accelerators, contributing to Galactic PeVatrons and potential multimessenger sources in our Galaxy.}

\maketitle

\section{Introduction}

Microquasars (MQs) are Galactic objects (also known as scaled down versions of quasars) with relativistic jets that are powered by the accretion process between a normal star and a compact object, either a neutron star or a stellar mass black hole \citep{Gallo2010LNP_5G,Ruston2012MNRAS_4R,Mirabel_2007CRPhy,chaty2005astro_6008C,paredes2002A_99P,Mirabel1999ARA_409M}. These are also known for their X-ray emission, and that could be either due to synchrotron emission of electrons in the jet or via external Compton upscattering of disk photons or photons of the companion star by the electron inside the jet \citep{Geor2002A_25G}. Radio flares from MQ GRS 1915+105 have been explained using synchrotron radiation model that provide valuable information about the ejecta, magnetic field and expansion properties of jets \citep{Atoyan1999MNRAS_53A}. Equipartition of energy densities provides magnetic field strength $\sim 0.2-0.3$ G during radio flares in GRS 1915+105. Recently, the Large High Altitude Air Shower Observatory (LHAASO) collaboration reported the observation of gamma-ray photons with energies above 100 TeV in our Galaxy. Five such spatial regions were linked to MQs: SS 433, V4641 Sgr, GRS 1915+105, MAXI J1820+070, and Cygnus X-1, respectively \citep{Lhaaso_2024arL}. These associations have been useful to probe MQ jet-cocoon structures as plausible Galactic PeVatrons \citep{Zhang2025a_93Z}. Charged particles can be accelerated to relativistic energies in these jets via shock acceleration \citep{Jones1991SSRv_259J}. These accelerated particles can be injected by these objects in the interstellar medium and might contribute towards the Galactic CR flux \citep{Ohira_2025MNRAS_34O,Heinz2002A_51H} and positron excess \citep{Gupta2014MNRAS_2G}. Individually, these objects are potential sources of gamma rays and neutrino flux \citep{Fang2024ApJ_35F,papa2021Galax_7P,Aharonian2006JP_8A,romero2003A_1R}. Hence, MQs can also accelerate cosmic ray nuclei (protons and heavy nuclei) and can be potential candidates for the neutrino emission \citep{Romero2005A_7R,zhang2010MN_68Z,Koljo2023MN_9K,Zhang2025a_93Z}. 

In the current sample of the TeV-PeV MQs, SS 433 was discovered by the High Altitude Water Cherenkov Observatory (HAWC), and photons up to 25 TeV were detected \citep{abeyse_2018Natur_A}. The High Energy Stereoscopic System (H.E.S.S.) also reported TeV gamma-ray observations from this source \citep{2024Sci_402H}. HAWC collaboration also discovered gamma rays in the range 1 to a few 100 TeV from V4641 Sgr \citep{alfaro2024ultra}. These gamma-ray observations have been well explained using leptonic radiation models. Persistent GeV gamma-ray emission from GRS 1915+10 can be explained by using cosmic ray proton interaction with the gas medium \citep{Mart2025ApJ_M}. However, in MAXI J1820 +070, photons in the energy range of 100 GeV have been explained by electrons of similar energies by scattering optical photons in the disc regions \citep{Rodi2021ApJ_21R,Abe2022MNRAS_36A}. Leptohadronic model has been described to study the origin of gamma ray emission above 100 GeV from Cygnus X-1 \citep{pepe_2015A_95P}. However, \citep{Zhang2014ApJ_43Z} found that synchrotron self-Compton and external Compton can explain the emission in the energy range from 100 MeV up to 1 TeV. The detection of gamma rays in the TeV–PeV range from microquasars is key to understanding their role as potential multimessenger sources. Motivated by the LHAASO observations of MQs, \citep{Kuze2025ApJ_K} have proposed the population of these objects as potential contributors to diffusion gamma ray emission above 100 TeV. On individual perspective \citep{carpio2025ar_550C} found that V4641 Sgr is the most promising objects for the next generation neutrino telescopes. Multiwavelength modeling of MQs also infers that protons can be accelerated beyond 1 PeV in the jets of SS433 \citep{Sudoh2020ApJ_146S}, and that makes them potential candidate for PeVatrons in our Galaxy. Utilizing the particle acceleration scenarios (magnetic reconnection, first and second order Fermi acceleration) in the jet of Cygnus X-3, it was shown that protons can be accelerated beyond PeV energies \citep{Kachel2025A_2K}. These recent progresses in UHE gamma-ray observations from microquasar jets provide us an opportunity to explore their role as an efficient accelerator of PeV cosmic ray protons and thereby a source of cosmic rays and Galactic PeVatrons. The stellar wind and the dense environment near the central binary may provide target material for proton-proton interactions. Thus, the possibility of jet-wind and jet-disk interaction allows for a hadronic contribution.

The origin of Galactic cosmic Rays remains a central problem in high-energy astrophysics \citep{Gabici2019IJ_022G}. Although supernova remnants (SNRs) have long been regarded as the prime candidates for accelerating CRs up to energies of approximately $10^{15}$ eV (below the 'knee' in the cosmic ray spectrum) \cite{blandford1978particle}, recent observations of PeVatron candidates require the identification of sources capable of accelerating particles to even higher energies. The detection of very-high-energy (VHE; $100\,\text{GeV} \lesssim E_\gamma \lesssim 100\,\text{TeV}$) and ultra-high-energy (UHE; $E \gtrsim 100 \,\text{TeV}$) gamma-rays along with TeV-PeV neutrinos directly probe the hadronic interactions, where accelerated protons collide with ambient matter, producing neutral pions that decay into gamma rays and charged pions that decay into neutrinos \cite{kelner2006energy}. Multi-messenger astronomy, combining observations of cosmic rays, photons, and neutrinos, is thereby the key to solving the mysteries of extreme astrophysical environments. Their powerful relativistic jets and dense surrounding environments provide natural conditions for particle acceleration and subsequent hadronic interactions. In this work, we focus on two Galactic microquasars, V4641 Sagittarii (V4641 Sgr) and SS 433, both of which exhibit relativistic jets and have recently been detected in TeV-PeV gamma-ray emission, making them potential candidates for studying hadronic emission processes. We apply a hadronic modeling framework to these microquasars, fitting the model with the recent VHE and UHE gamma-ray detections and predicting their corresponding neutrino fluxes. We thus assess them as Galactic PeVatrons and multimessenger sources - sources of cosmic rays, gamma-rays and neutrinos.

\section{Secondary Gamma-Rays and Neutrino Flux:  $P-P$ Interactions}
\label{model}

Following the hadronic jet framework introduced in \cite{romero2003hadronic}, we present a hadronic model for the production of gamma rays from microquasars, valid for $E_{\gamma} \geq 1 \,\text{GeV}$. They have used a jet-disk coupling scenario to estimate the amount of matter density available in the jet \citep{Falcke1995A_65F}. If $\dot{M}_{\text{disk}}$ is the mass loss rate of the disk around the compact object then the total jet power is $Q_j = q_j \dot{M_{\rm disk}} c^2$, where $q_j$ is the efficiency of the accretion process and its value lies in between $10^{-1}$ to $10^{-3}$ \citep{romero2003hadronic}.  The amount of gas density in the jet is determined using the stellar wind of the companion star and assuming a proton only wind environment, we have $n(z) = \dot{M_{\ast}} \left(1- r_{\ast}/\sqrt{\left(z^2+a^2\right)}\right)^{-\zeta}/4 \pi m_p v_{\infty} (z^2 + a^2)$. Here $\dot{M}_{\ast}$ is the mass loss rate of the companion star and $r_{\ast}$ is the companion star radius. Direction $z$ is along the jet axis, which is normal to the orbital radius $a$, and $v_{\infty}$ is the terminal wind speed and wind velocity index $\zeta \sim 1$ for massive stars.

The distribution of cosmic ray protons in the relativistic jet is described by a power law $N'_p(E'_p) = K_p E'^{-\alpha_p}_p$, where $\alpha_p$ is the spectral index. Normalization of this spectrum $K_p$ is estimated using the gas that is available at the jet base $z_0$ \citep{romero2003hadronic}. The density of gas at $z_0$, the base of the jet, is given by $n_0= Q_j/({\pi c^3R_0^2m_p})$, where $m_p$ is the rest mass of the proton and $R_0$ is the radius of the jet at $z_0$. 
\noindent
The resultant pion decay gamma-ray luminosity at energies \( E_\gamma > 1\,\text{GeV} \), in \( \theta \)  direction with respect to the jet axis, is given by \cite{romero2003hadronic}
\begin{equation}
\begin{aligned}
L^{\pi^0}_\gamma(E_\gamma, \theta) &\approx E_\gamma^2\frac{q_j z_0^{\epsilon (n - 2)} Z^{(\alpha_p)}_{p \rightarrow \pi^0}}{2\pi m_p^2 v_\infty}  \frac{(\alpha_p - 1)}{\alpha_p}\left(E'^{\text{min}}_p\right)^{\alpha_p - 1}\times \dot{M}_\ast \, \dot{M}_{\text{disk}}\times \sigma_{pp}(10 E_\gamma) \\
&\quad \times\frac{ \Gamma^{-\alpha_p + 1} \left(E_\gamma{-\beta} \sqrt{E_\gamma^2 - m_p^2 c^4} \cos \theta\right)^{-\alpha_p} } {\,
\left[
( \sin^2 \theta + \Gamma^2 \left(\cos\theta -  \frac{\beta E_\gamma}{\sqrt{E_\gamma^2 - m_p^2 c^4}} \right)^2
\right]^\frac{1}{2}}\\
&\quad \times  \int_{z_0}^{\infty} \frac{z^{\epsilon(2-n)}}{(z^2 + a^2)} \left(1 - \frac{r_\ast}{\sqrt{z^2 + a^2}}\right)^{-\zeta} \, dz,
\label{eq:lm}
\end{aligned}
\end{equation}

\noindent 
where $\Gamma$ is the Lorentz factor of the jet. $z_0 \ge 100$ km, is the height of the jet base. The radius of the jet having lateral expansion is parametrized using the relation $R(z) = \xi z^{\epsilon}$, and $\epsilon \le 1$. More details on the derivation of the above expression are described in \citep{romero2003hadronic}. 

Gamma rays can be produced through both leptonic and hadronic processes, making the detection of gamma-ray emissions alone insufficient for definitively identifying the source type as leptonic or hadronic. To distinguish between leptonic and hadronic sources, additional observational evidence is needed. The detection of neutrinos from a source is one of such critical evidence, confirming the acceleration of hadronic particles in the source. Neutrinos can be produced in astrophysical environments when the accelerated hadronic particles, such as protons or heavy nuclei, interact with a target proton or photon distribution. In this calculation of the neutrino flux, we will focus solely on the proton-proton interactions. As a typical MQ is capable of accelerating protons and nuclei to extremely high energies, which will then interact with the target protons present in the stellar wind of the companion star. In contrast, proton-gamma interactions require the presence of a strong photon field and/or ultra-relativistic accelerated hadrons, conditions that can only be found in ultra-luminous, photon-rich MQs. Currently, no observational evidence of neutrinos from V4641 Sgr and SS433 has been detected. However, we tried to estimate the neutrino flux from our best-fitted gamma-ray flux of the sources following the formalism defined in \cite{ahlers2014probing}. The predicted all-flavor neutrino flux is given by,
\begin{equation}
    E_{\nu}^{2} \, \Phi_{\nu}(E_{\nu}) \approx \frac{3}{2} \, E_{\gamma}^{2} \, \Phi_{\gamma}(E_{\gamma}),
    \label{eq:nu}
\end{equation}
where $E_{\nu} \approx E_{\gamma}/2$. $\Phi_{\nu}(E_{\nu})$ is the differential neutrino flux and $\Phi_{\gamma}(E_{\gamma})$ is the differential gamma-ray flux. The factor 3/2 accounts for the energy distribution among decay products in pion decays, assuming a charged-to-neutral pion production ratio of approximately 2:1 in $P-P$ collisions. Using the relation \eqref{eq:nu}, relating gamma-ray and neutrino flux in hadronic interactions, we compute the spectral energy distribution of all-flavor neutrinos at Earth corresponding to our best-fit models. The neutrino flux contributions from V4641 Sgr are illustrated in fig. \ref{fig:sgr}, along with the best-fit gamma-ray flux, observed data points, and the sensitivity of the KM3NeT and TRIDENT neutrino detectors. Similarly, the contributions for SS433 are shown in fig. \ref{fig:ss433}.


\section{Spectral Energy Distribution: VHE to UHE Gamma-Rays}
\label{sed}
The modeled gamma-ray flux will be fitted to the observed VHE and UHE gamma-ray data from HAWC, HESS and LHAASO using the Bayesian Markov Chain Monte Carlo (MCMC) method. The free parameters were the jet power fraction carried by hadrons ($q_j$), the proton spectral index ($\alpha$), the maximum proton energy ($E'^{\max}_p$) and the bulk Lorentz factor ($\Gamma$). The likelihood accounted for asymmetric errors on the observed flux points. We adopted flat priors which were imposed within physically motivated ranges $1.8 \leq \alpha \leq 3$, $-6 \leq \ln q_j \leq -1$, $0.1 \leq E'^{\max}_p \leq 10$ PeV and $1 \leq \Gamma \leq 10$ \cite{falcke1996galactic,pepe2015lepto,escobar2021cosmic,escobar2022highly}. Below, we describe the gamma-ray observations of V4641 Sgr and SS433 and their multimessenger modeling.

\subsection{V4641 Sgr}

V4641 Sagittarii (V4641 Sgr) is located at a distance of about 6.6$~\mathrm{kpc}$ \cite{orosz2001black} from the Earth. It is a binary system composed of a $\sim6.4\,M_\odot$ black hole \cite{macdonald2014black, andrae2018gaia} and a $\sim3.0\,M_{\odot}$ B9III-type subgiant companion star with radius $5.3\,R_\odot$ \cite{macdonald2014black} and weak mass loss rate $\sim10^{-9}~ {M_{\odot}/{\rm yr}}$ \cite{vink2000new,lamers1999introduction}. It was first identified as an X-ray source in 1999 with a subsequent radio outburst observed by the Very Large Array (VLA), revealing superluminal motion of ejecta and thus confirmed as a microquasar \cite{hjellming2000light}. They resolved a bright, jet-like structure of length $\approx0.25''$. V4641 Sgr is particularly notable for its episodes of super-Eddington mass accretion onto the black hole $\sim10^{-7}~ {M_{\odot}/{\rm yr}}$ \cite{revnivtsev2002} and its highly relativistic jet which is among the fastest superluminal jets in the Milky Way \cite{hjellming2000light}. The jet is closely aligned to our line of sight with jet angle $<12^\circ$ \cite{orosz2001black}. Recently its X-ray observation\cite{shaw2022high} suggests the presence of accretion disk wind with spherical symmetry. High-energy gamma-ray observations have recently revealed V4641 Sgr as a potential extreme particle accelerator and Galactic PeVatron \cite{alfaro2024ultra,neronov2025multimessenger,suzuki2025detection}. The LHAASO and the HAWC observatory both have detected emission extending beyond 100~TeV. HAWC \cite{alfaro2024ultra} reported a maximum statistical significance of $8.8\sigma$ above 1~TeV and $5.2\sigma$ above 100~TeV from 2014-2022 monitoring. Their gamma-ray morphology appears extended, exhibiting a jet-like structure spanning $\sim 1^\circ$-$2^\circ$ from the source position, suggesting existence of a large-scale steady jet. LHAASO \cite{lhaaso2024ultrahigh} have reported the detection of an elongated source from this microquasar with gamma ray energy extending up to 800 TeV.

\begin{figure}[h!]
    \centering
    \begin{subfigure}[b]{0.55\textwidth}
        \centering
    \includegraphics[width=\textwidth]{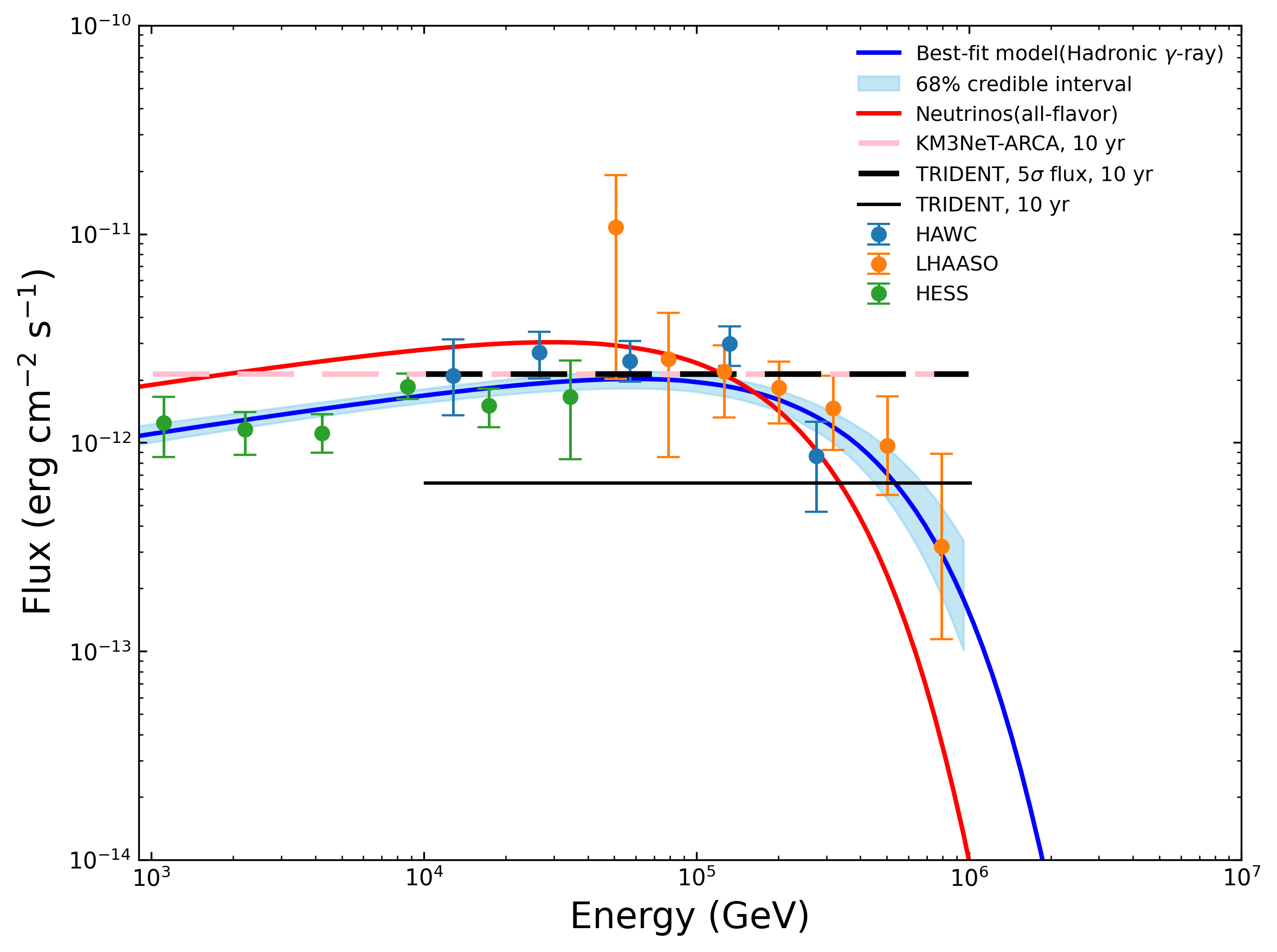}
    \caption{}
    \end{subfigure}
    \hfill
    \begin{subfigure}[b]{0.48\textwidth}
        \centering
        \includegraphics[width=\textwidth]{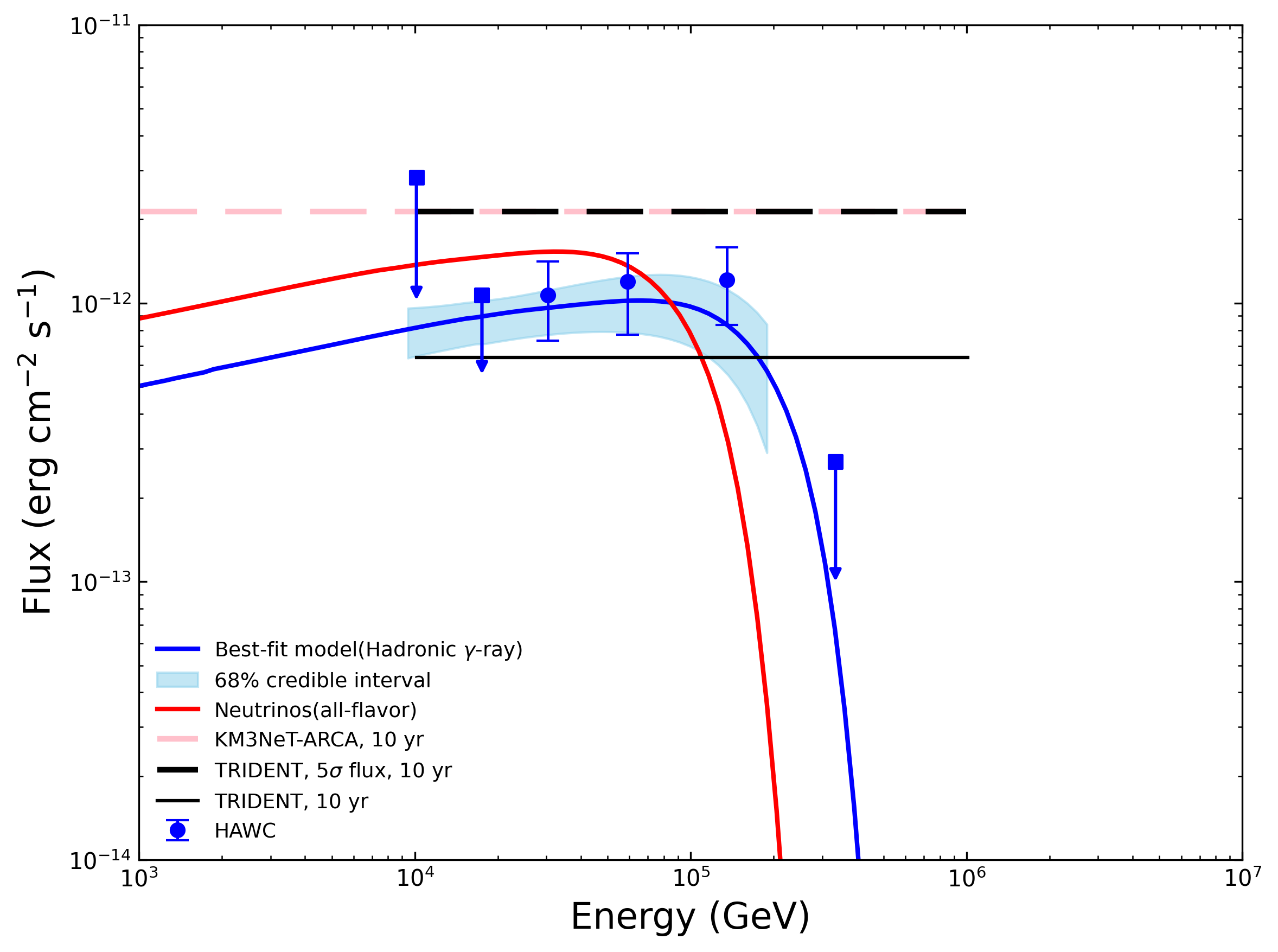}
        \caption{}
    \end{subfigure}
    \hfill
    \begin{subfigure}[b]{0.48\textwidth}
        \centering
        \includegraphics[width=\textwidth]{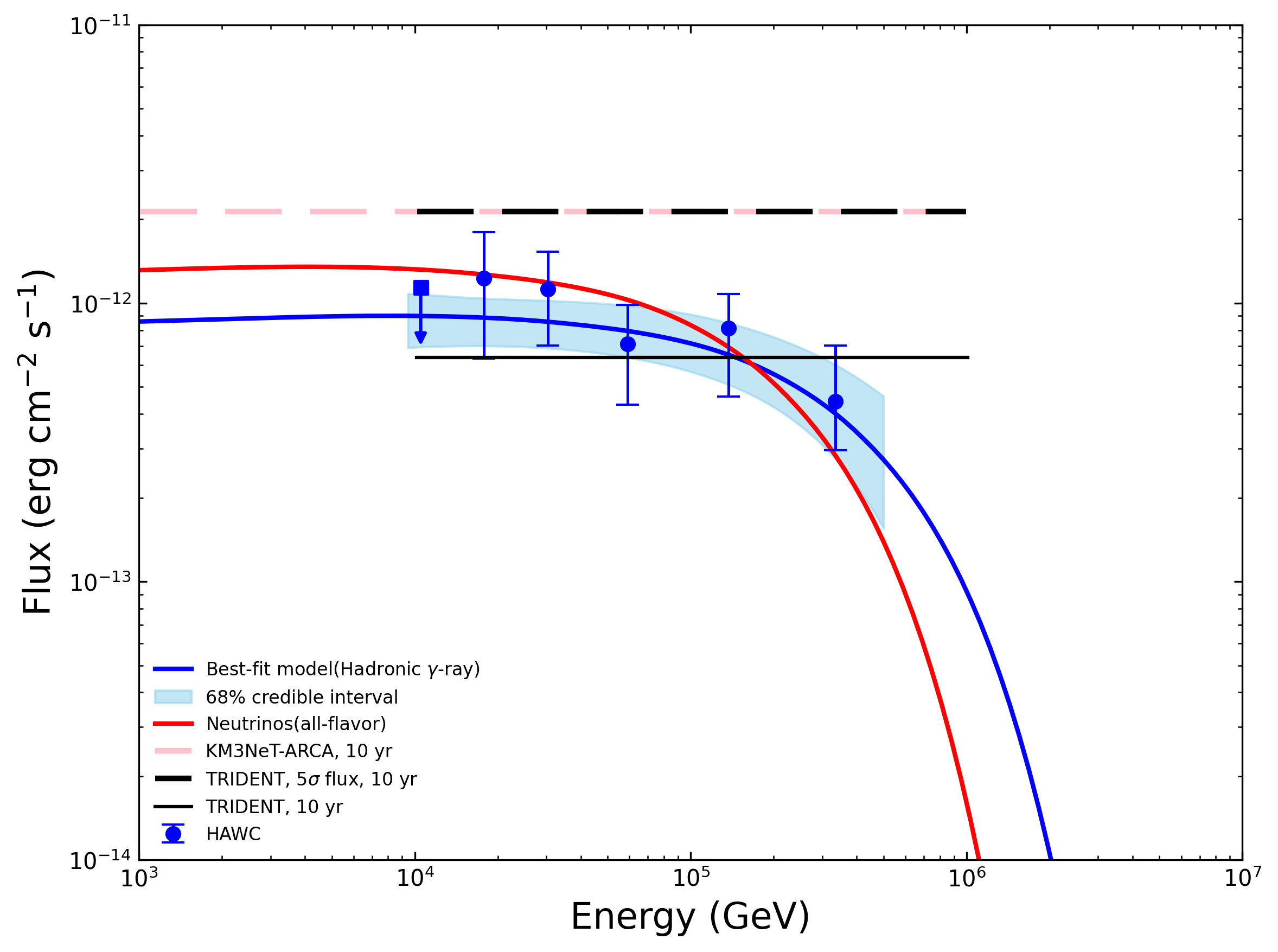}
        \caption{}
    \end{subfigure}

    \caption{Spectral energy distribution of VHE and UHE gamma-ray emission from the hadronic model for the single extended source (a), northern site (b) and southern site (c) of HAWC ROI of the microquasar V4641 Sgr. The blue line represents the best-fit hadronic model predicted gamma-ray flux. The light blue band indicates the corresponding $68\%$ (equivalent to $1\sigma$) credible intervals derived from the MCMC posterior samples. The solid red line shows the predicted all-flavor neutrino flux corresponds to the hadronic gamma-ray flux. Blue, orange and green data points with error bars are the observed gamma-ray data reported by HAWC \cite{alfaro2024ultra}, LHAASO \cite{Lhaaso_2024arL} and HESS \cite{neronov2025multimessenger} respectively. The Dashed green and black lines indicate neutrino flux sensitivity level by KM3NeT-ARCA, 10 years and TRIDENT $5\sigma$ discovery for 10 years respectively, and the solid black line for 10 years of TRIDENT data (extracted from \cite{neronov2025multimessenger}). Both the sensitivity lines have been plotted by dashed lines with gaps for visual distinction.}
    \label{fig:sgr}
\end{figure}

\subsection{SS433}
SS433 is located at a distance of about 5.5$~\mathrm{kpc}$ \cite{ghisellini1985inhomogeneous,blundell2004symmetry} from the Earth. It is composed of a compact object, most likely a black hole of mass $15\, M_\odot$ \cite{bowler2018ss,cherepashchuk2019mass} and an A7-type supergiant star with radius $50\, R_\odot$ \cite{hillwig2008spectroscopic}. The star overflows its Roche lobe, resulting in a supercritical accretion flow onto the black hole at $\sim 10^{-4}~ {M_{\odot}/{\rm yr}}$ \cite{ahnen2018constraints,begelman2006nature}. The lower limit on mass loss rate from the star has been reported as $\sim 7\times10^{-6}~ {M_{\odot}/{\rm yr}}$ \cite{cherepashchuk2021discovery}. The black hole accretes matter from the star and gives rise to a pair of jets moving in opposite directions with approximately one quarter of the speed of light and at almost perpendicular to our line of sight \cite{fabrika1997supercritical}. The jet precesses with a half-opening angle of
$20^\circ$ \cite{fabrika2004astrophys}. Radio and optical observations show precessing relativistic jets from the black hole extend up to $10^{-3}$ pc and 0.1 pc respectively. These jets inflate lobes, referred to as east and west lobe \cite{geldzahler1980continuum,safi1997rosat}. HAWC \cite{alfaro2024spectral} reported the detection of two point-like sources, coinciding with the known east and west lobes of SS 433. They found gamma ray emission from the east lobe reaching maximum energy of 56 TeV and west lobe with maximum energy of 123 TeV, but no evidence of extended emission for these lobes. LHAASO \cite{lhaaso2024ultrahigh} has detected extended UHE gamma ray emission(100-400 TeV) from the central region of SS 433. They also reported 25-100 TeV gamma ray emission from the two resolved point-like sources in the east and west lobes. These TeV gamma-rays can not be explained by purely leptonic model due to the Klein-Nishina effect \cite{cristofari2021hunt}. Therefore, it strongly indicates hadronic contribution where protons accelerated in PeV energies, interact with ambient matter to produce observed TeV gamma-rays.

\begin{figure}[h!]
    \begin{subfigure}[b]{0.55\textwidth}
        \centering
        \includegraphics[width=\textwidth]{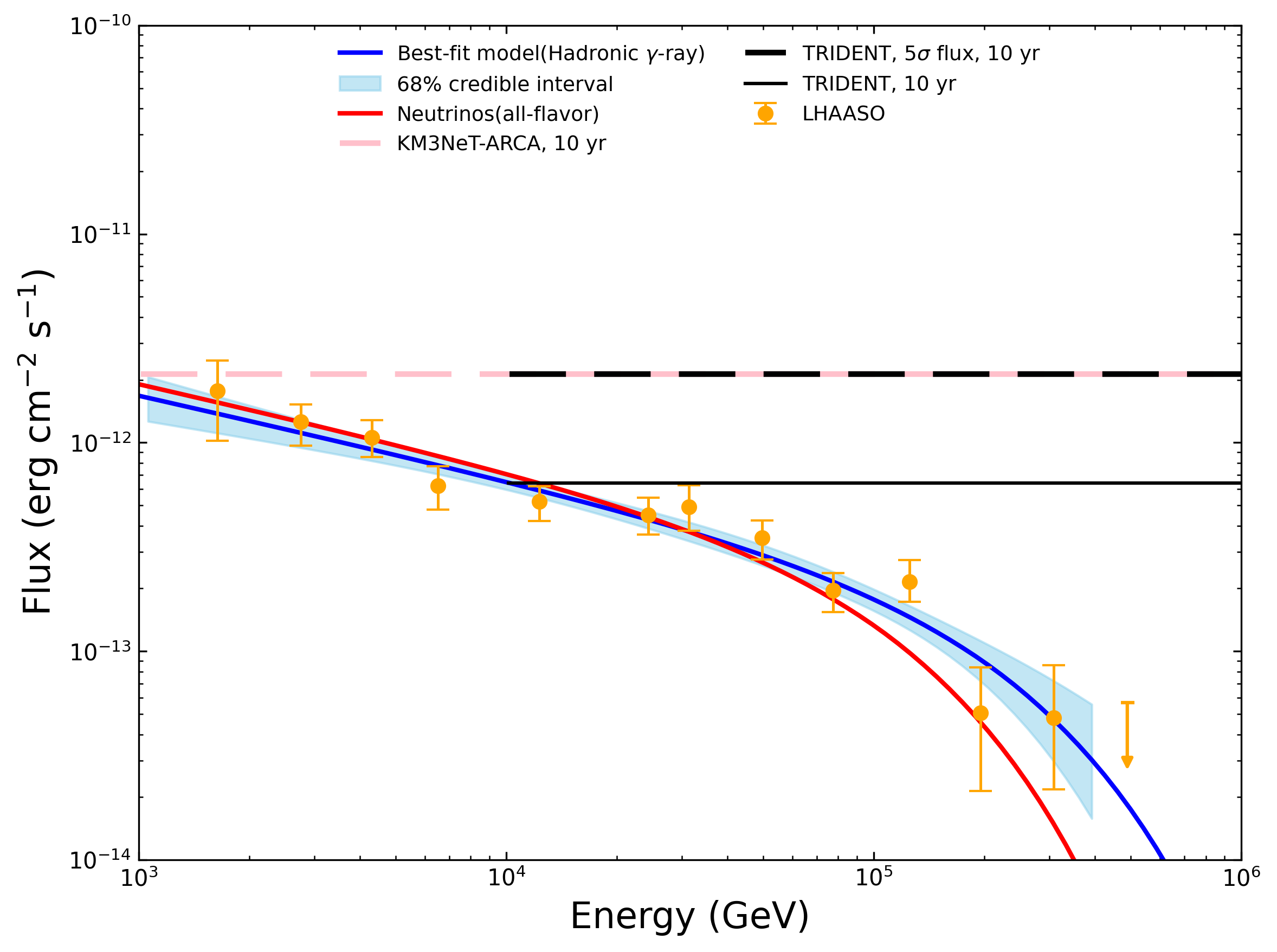}
        \caption{}
    \end{subfigure}
    \hfill
    \begin{subfigure}[b]{0.48\textwidth}
        \centering
        \includegraphics[width=\textwidth]{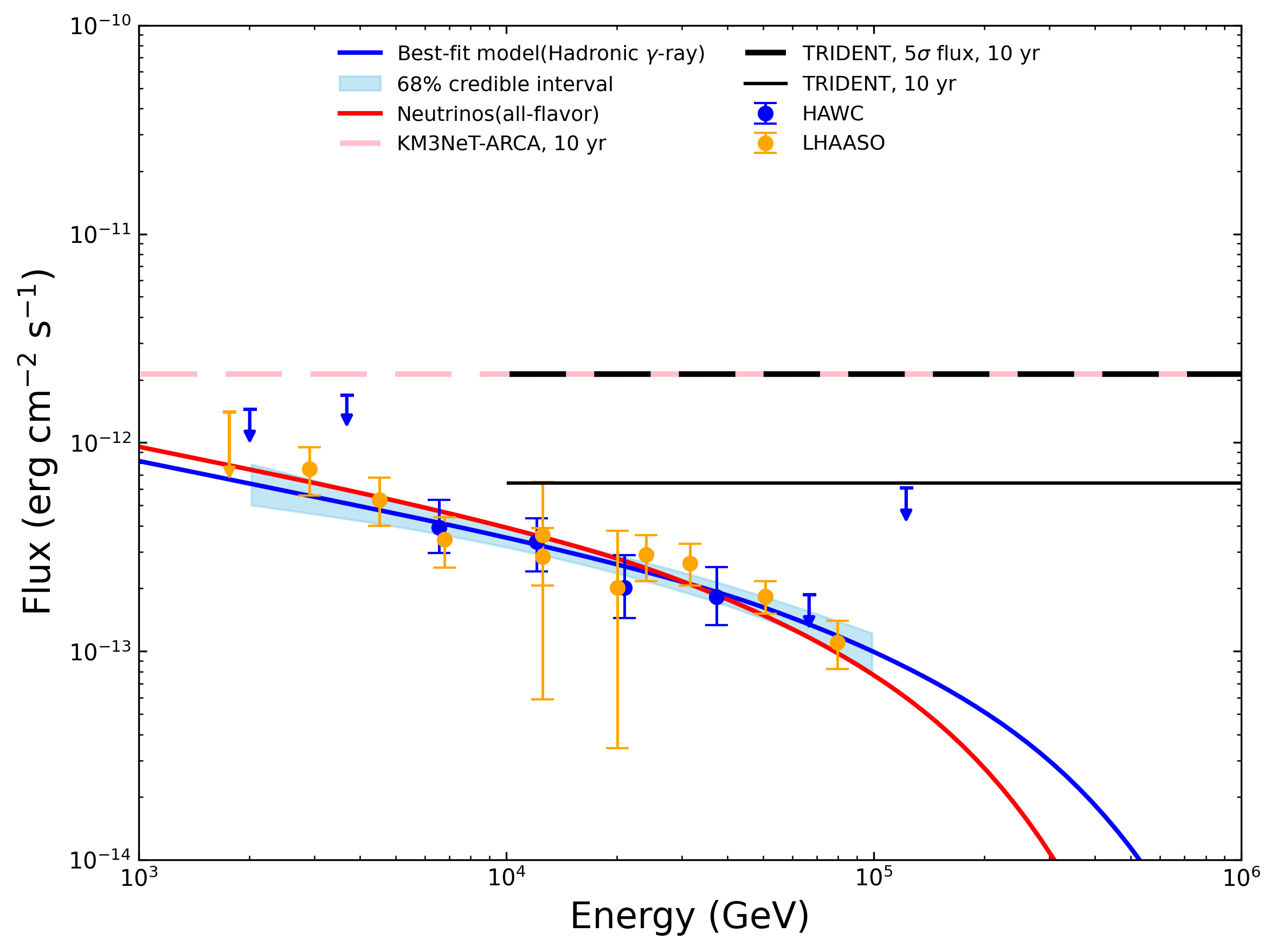}
        \caption{}
    \end{subfigure}
    \hfill
    \begin{subfigure}[b]{0.48\textwidth}
        \centering
        \includegraphics[width=\textwidth]{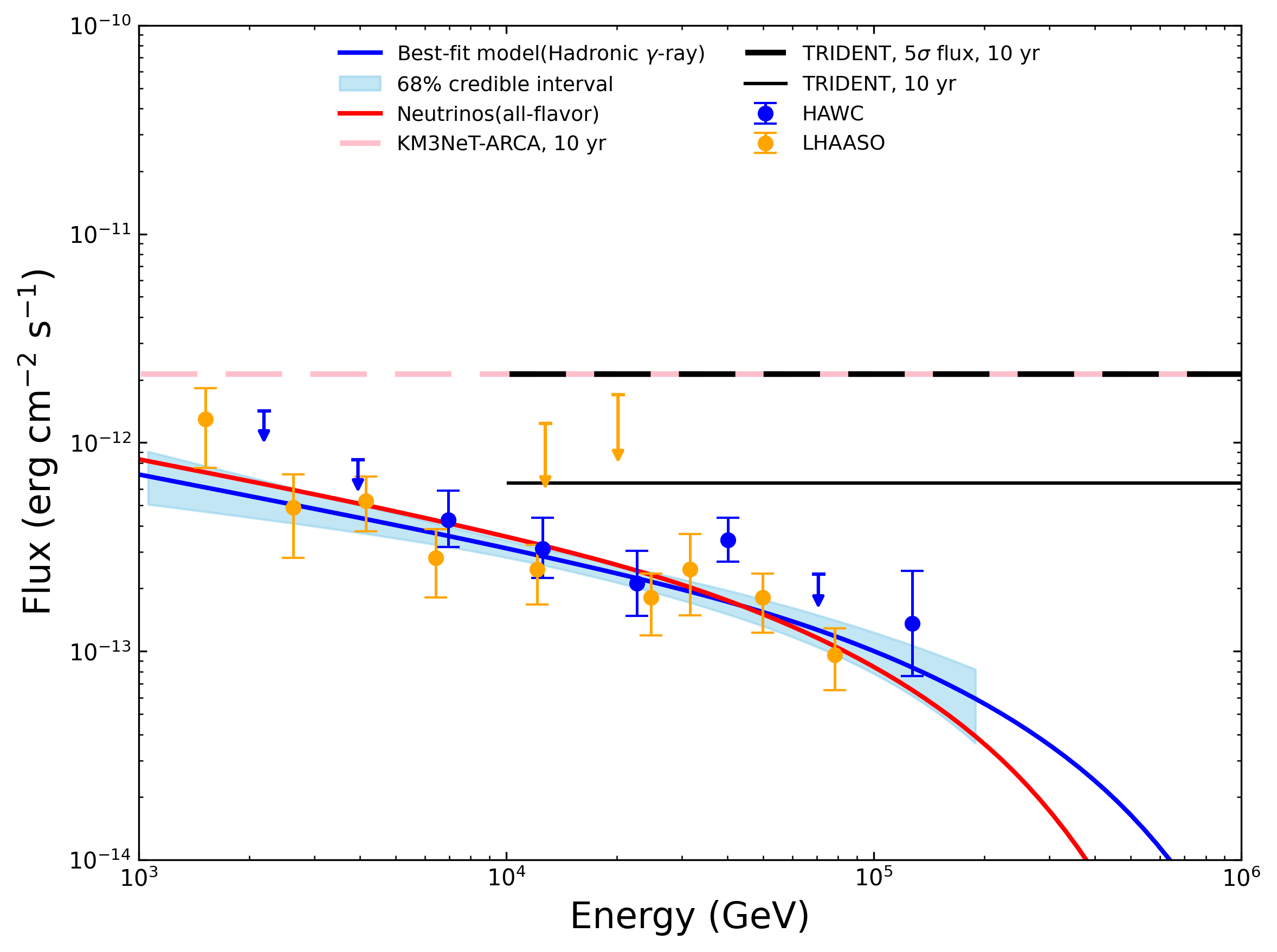}
        \caption{}
    \end{subfigure}
    \caption{Spectral energy distribution of VHE and UHE gamma-ray emission from the hadronic model for the east lobe(a), west lobe(b) and the total summed up emission(d) of SS 433. The blue line represents the best-fit hadronic model predicted gamma-ray flux. The light blue band indicates the corresponding $68\%$ (equivalent to $1\sigma$) credible intervals derived from the MCMC posterior samples. The solid red line shows predicted all-flavor neutrino flux corresponds to the modeled gamma-ray flux. Blue, orange and green data points with error bars are the observed gamma-ray data reported by HAWC\cite{alfaro2024spectral}, LHAASO\cite{Lhaaso_2024arL} and HESS\cite{neronov2025multimessenger} respectively. The Dashed green and black lines indicate neutrino flux sensitivity level by KM3NeT-ARCA, 10 years and TRIDENT $5\sigma$ discovery for 10 years respectively, and the solid black line for 10 years of TRIDENT data (extracted from \cite{neronov2025multimessenger}). Both the sensitivity lines have been plotted by dashed lines with gaps for visual distinction.}
    \label{fig:ss433}
\end{figure}

\section{Population Modeling and Galactic Diffuse Gamma Ray emission}
\label{diff gamma}
The LHAASO collaboration recently published results on the truly diffuse \(\gamma\)-ray emissions above 10 TeV \citep{Cao2023PhRvL_01C}, focusing on two regions within the Galaxy: the inner region (\(15^{\circ} < l < 125^{\circ}, |b| \le 5^{\circ}\)) and the outer region (\(125^{\circ} < l < 235^{\circ}, |b| \le 5^{\circ}\)). They masked out active source regions listed in the TeVCat \cite{2008ICRC....3.1341W} and LHAASO source catalog \cite{2024ApJS..271...25C}. The study found that the diffuse emissions in the inner and outer galaxy are, respectively, three and two times higher than the contributions from the interaction of diffuse GCR observed near Earth with the distribution of gas density in our Galaxy. This suggests the presence of some unresolved source contributions in the diffuse \(\gamma\)-ray emission.

\begin{figure}[h!]
    \hfill
    \begin{subfigure}[b]{0.48\textwidth}
        \centering
    \includegraphics[width=\textwidth]{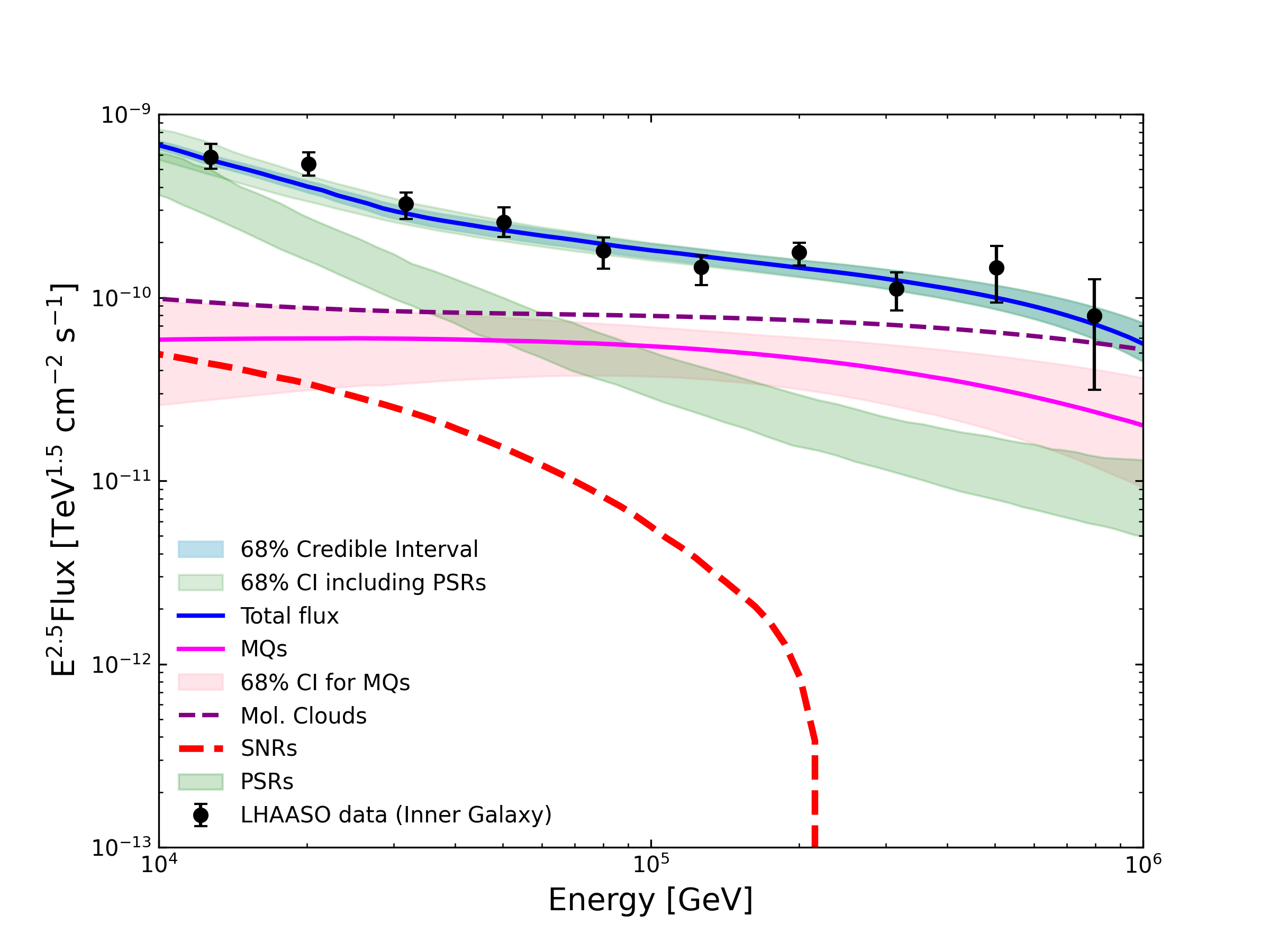}
    \caption{}
    \end{subfigure}
    \hfill
    \begin{subfigure}[b]{0.5\textwidth}
        \centering
        \includegraphics[width=\textwidth]{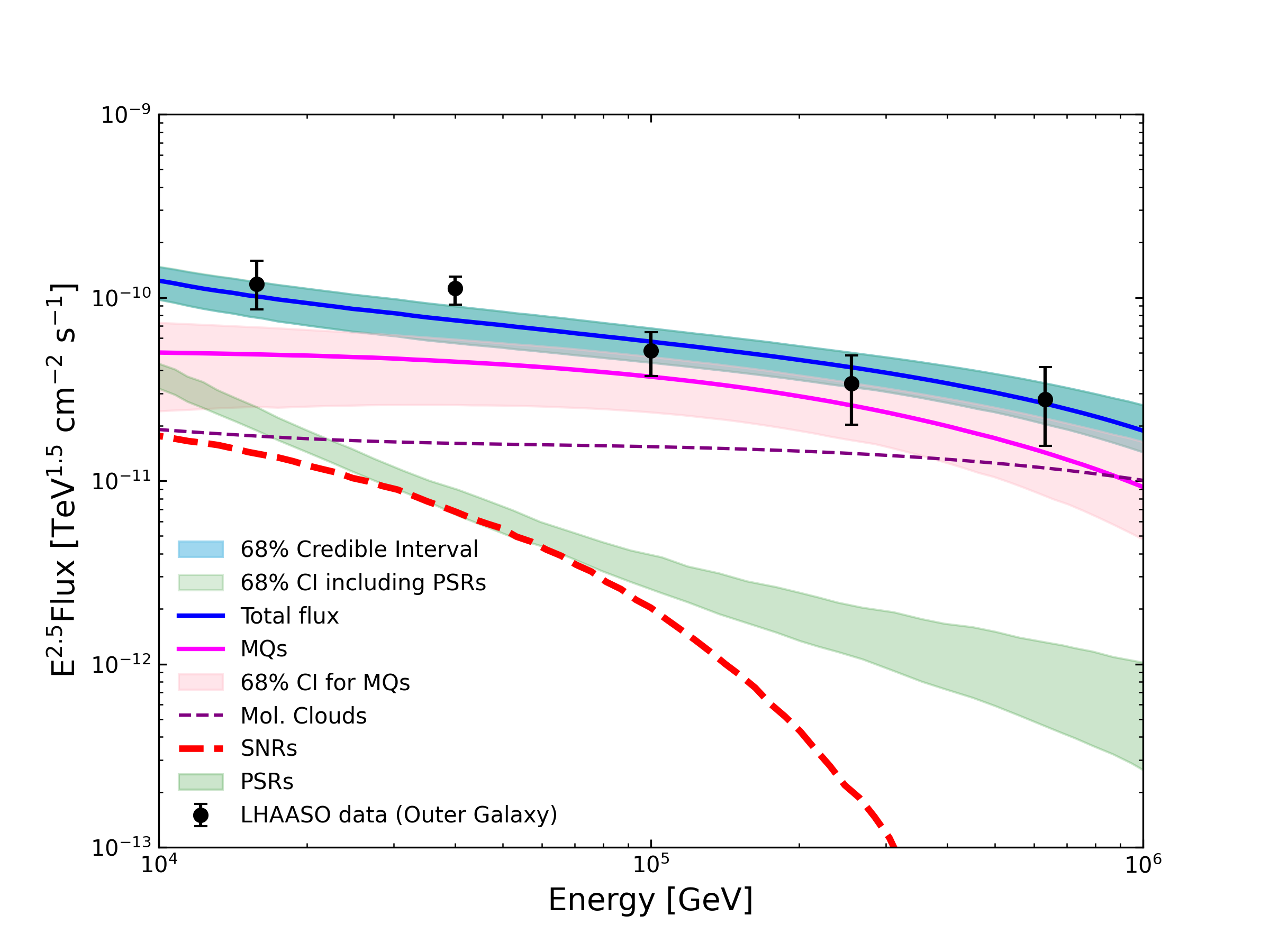}
        \caption{}
    \end{subfigure}

    \caption{Gamma ray flux contribution from a population of Galactic microquasars (MQs), Molecular Clouds (MCs), Pulsars (PSRs) and Supernova Remnants (SNRs) for the Galactic diffuse gamma ray emission of inner (a)  and outer galaxy (b). The black points with error bars are the LHAASO observed diffuse gamma ray data for inner and outer galaxy respectively (extracted from \cite{Cao2023PhRvL_01C} and increased by 61\% and 2\% for inner and outer galaxy respectively). The pink solid line with lightpink shaded region represents the population modeled gamma-ray flux of MQs for $E'^{\max}_p$ up to 10 PeV. The blue solid line is the total flux contribution from MQS, MCs, PSRs and SNRs with corresponding shaded region indicating the $68\%$ credible intervals derived from 100 simulations. The red dashed line and green region show contributions from SNRs and PSRs (extracted from \cite{kaci2025microquasars}). The purple dashed line represents the contribution from MCs (taken from \cite{Roy24}).}
    \label{Fig:diff}
\end{figure}

Possible sources for these emissions include Galactic binary, microquasars (MQs), supernova remnants (SNRs), young massive stellar clusters (YMSCs), and pulsar wind nebulae (PWNe), among others \cite[and references therein]{Bykov26, kaci2025microquasars, Cristofari21, Aharonian19, Dekker24, Martin22}. In the past, several authors have tried to explain the diffuse emissions by considering contributions from YMSCs \cite{Menchiari25}, MQs \cite{kaci2025microquasars}, and interactions between dense giant molecular clouds (GMCs) with a non-uniform GCR distribution \cite{Roy24}. However, in this section, we revisited this problem to account for the diffuse $\gamma$-ray emission observed by LHAASO, with particular emphasis on the hadronic contributions from MQs within the aforementioned formal framework. In addition, we incorporate the potential contributions from pulsars (PSRs), supernova remnants (SNRs), and molecular clouds (MCs) from \cite{kaci2025microquasars, Roy24}. The LHAASO analysis found the diffuse gamma-ray emission in the Regions of Interest(ROIs) with applied masking, and for the unmasked diffuse background, the fluxes will be higher by 61\% (inner region) and 2\% (outer region). Therefore, we increased the measured fluxes by these respective factors and re-elevated the observation using the solid angle of the ROIs for the comparison with our modeling. We created a synthetic population of Galactic MQs whose intrinsic properties and surrounding environmental conditions lie within physically motivated ranges and are consistent with our best-fit solutions for V4641 Sgr and SS433. The model parameters including stellar mass loss rate ($\dot{M}_\ast$), black hole accretion rate ($\dot{M}_{\mathrm{disk}}$), source distance to Earth ($d$), jet bulk Lorentz factor ($\Gamma$), injection proton spectral index ($\alpha_p$), jet-disk coupling constant ($q_j$) and maximum proton energy ($E'^{\max}_p$) are sampled within the defined prior ranges, $1.05\times10^{-9} \leq \dot{M}_\ast \leq 1.05\times10^{-4}$ $\,M_\odot\,\mathrm{yr}^{-1}$, $1.05\times10^{-9} \leq \dot{M}_{\mathrm{disk}} \leq 1.05\times10^{-4}$ $\,M_\odot\,\mathrm{yr}^{-1}$, $1.5 \leq d \leq 10$ kpc, $1 \leq \Gamma \leq 10$, $1.8 \leq \alpha \leq 2.6$, $-6 \leq \ln q_j \leq -1$ and $1 \leq E'^{\max}_p \leq 10$ PeV. We have included an exponential cutoff $\exp[-(E_\gamma / (0.1\, E'^{\max}_p)^{0.5}]$ \cite{lefa2012spectral}. The contribution of diffuse gamma-ray flux from a population of MQs for the inner and outer Galaxy is shown in fig. \ref{Fig:diff}. Based on this calculation, we have found that after accounting for the flux contributions from other relevant source classes (e.g., MCs, PSRs, SNRs), approximately $\sim$ 14 MQs with $E'^{\max}_p$ up to 10 PeV are sufficient to explain the observed diffuse \(\gamma\)-ray emission in both the inner and outer Galaxy. This result is consistent with the conclusions of \citep{kaci2025microquasars}, who inferred that a population of \(\sim10\) MQs can adequately account for the observed diffuse \(\gamma\)-ray flux. We found the best fit proton spectra 2.43 and 2.50; the maximum proton energy 3.19 PeV and 2.15 PeV to explain the diffuse TeV gamma-ray data from inner and outer galactic regions respectively. 




\section{Results and Discussion}

Here we present the results of the MCMC fits and further findings for V4641 Sgr and SS 433. In \cref{fig:sgr} and \cref{fig:ss433}, we present the modeled spectral energy distribution (SED) of gamma-ray along with the predicted all-flavor neutrino flux from hadronic interaction and compare with the observational data for the microquasars V4641 Sgr and SS 433 respectively. The TeV gamma ray emission detected from V4641 Sgr and SS 433 are well described by the hadronic model with hard proton index $(1.84-2.44)$ and their maximum energies reaching $ 2.1$ PeV to $4.5$ PeV, consistent with the PeVatron nature suggested by LHAASO. The jet-disk coupling, which relates $\dot{M}_{\mathrm{disk}}$ and $q_j$, as discussed in section \ref{model} is also justified by our results. For V4641 Sgr, we found higher $q_j$ values $\sim(1-6)\times10^{-3}$ for the given low value of $\dot{M}_{\mathrm{disk}}$, whereas in case of SS 433 we found relatively lower $q_j$ values $\sim1.1\times10^{-5}- 5.8\times10^{-6}$ for the given high $\dot{M}_{\mathrm{disk}}$ value. 

\begin{table*}[t]
\centering
\resizebox{\textwidth}{!}{
\begin{tabular}{lccc}
\toprule
\textbf{Model parameters} & \textbf{V4641 Sgr} & \textbf{SS 433} \\
\midrule
\multicolumn{3}{l}{\textbf{From Past Observations}} \\[2pt]
Companion star radius: $r_\ast$(cm) & $5.3\,R_\odot$\cite{macdonald2014black} & $50\,R_\odot$\cite{hillwig2008spectroscopic}\\
Orbital radius: $a$(cm) & $1.22\times10^{12}$\cite{salvesen2020origin} & $7.4\times10^{12}$\cite{begelman2006nature}\\
Black hole mass: ${M}_{\mathrm{bh}}$ ($\mathrm{gm}$) & $6.4\,M_\odot$ \cite{macdonald2014black,andrae2018gaia} & $15\,M_\odot$\cite{bowler2018ss,cherepashchuk2019mass}\\
Stellar mass loss rate: $\dot{M}_\ast$($\,M_\odot\,\mathrm{yr}^{-1}$) & $10^{-9}$\cite{vink2000new,lamers1999introduction} & $7\times10^{-6}$\cite{cherepashchuk2021discovery}\\
Black hole accretion rate: $\dot{M}_{\mathrm{disk}}$($\,M_\odot\,\mathrm{yr}^{-1}$) & $10^{-7}$\cite{done2007modelling,revnivtsev2002,hjellming2000light} & $10^{-4}$\cite{king2000evolutionary,fabrika2004astrophys}\\
Terminal wind velocity: $v_\infty$($\,\mathrm{km\,s}^{-1}$) & 1600\cite{munoz2018low,lamers1999introduction} & 1500\cite{perez2009inflow,lamers1999introduction}\\
Jet precession angle: $\theta$($^\circ$) & 12\cite{orosz2001black} & 20\cite{monceau2015ss433}\\
Distance to Earth: $d$($~\mathrm{kpc}$) & 6.2\cite{macdonald2014black} & 5.5\cite{ghisellini1985inhomogeneous}\\[5pt]
\multicolumn{3}{l}{\textbf{Assumed Parameters}} \\[2pt]
Type of jet: $\epsilon$ & 1 & 1\\
Jet expansion index: $n$ & 2 & 2\\
Base of the jet: $z_0$($\,\mathrm{km}$) & 50$R_g^1$ & 50$R_g^1$\\
Wind velocity index: $\zeta$ & 1 & 1\\
Minimum proton energy: $E'^{\min}_p$(GeV) & 2 & 2\\[5pt]
\multicolumn{3}{l}{\textbf{Fitted Parameters}} \\[2pt]
Injection proton spectral index: $\alpha_p$ & $1.84^{+0.05}_{-0.03}$ (Extended) & $2.44^{+0.11}_{-0.12}$ (Total)\\[2pt]
& $1.84^{+0.18}_{-0.10}$ (Northern site) & $2.39^{+0.11}_{-0.14}$ (East lobe)\\[2pt]
& $2.00^{+0.14}_{-0.12}$ (Southern site) & $2.38^{+0.11}_{-0.13}$ (West lobe)\\[5pt]

Jet-disk coupling constant: $\log_{10}q_j$ & $-2.84^{+0.19}_{-0.11}$ (Extended) &
$-4.95^{+0.46}_{-0.51}$ (Total)\\[2pt]
& $-2.99^{+0.86}_{-0.49}$ (Northern site) & $-5.12^{+0.49}_{-0.65}$ (East lobe)\\[2pt]
& $-2.22^{+0.63}_{-0.52}$ (Southern site) & $-5.24^{+0.49}_{-0.59}$ (West lobe)\\[5pt]

Maximum proton energy: $E'^{\max}_p$(PeV) & $3.03^{+1.10}_{-0.50}$ (Extended) & $2.41^{+3.81}_{-0.94}$ (Total) \\[2pt]
& $2.13^{+0.50}_{-0.52}$ (Northern site) & $2.56^{+3.83}_{-1.54}$ (East lobe)\\[2pt]
& $4.47^{+3.26}_{-1.95}$ (Southern site) & $3.36^{+3.77}_{-1.98}$ (West lobe)\\[5pt]

\multicolumn{3}{l}{\textbf{Derived Parameters}}\\[2pt]
Gas density at $z_0$: $n_0$($\rm cm^{-3}$) & $1.90\times10^{7}$ & $5.08\times10^{9}$\\
Constant in proton spectrum at $z_0$: $K_0$($\mathrm{cm^{-3}\,GeV^{\alpha_p-1}}$) & $4.06\times10^{13}$ & $1.50\times10^{14}$\\
\bottomrule
$^1$ $R_g= GM_{bh}/c^2$ is the gravitational radius
\end{tabular}
}
\caption{Hadronic gamma-ray emission model parameters for MQs V4641 Sgr and SS 433.}
\label{tab:model_parameters}
\end{table*}

For V4641 Sgr we performed three spectral fits. The first is an extended emission fit, where we consider a joint analysis using HAWC, LHAASO and HESS measurements of the source, thereby constraining the global proton spectrum responsible for the observed TeV emission. The second and third are the northern and southern region fit using the HAWC ROI data corresponding to the northern and southern extension of the source, respectively. This separate fit allows us to probe potential spectral variations within the extended source.

\noindent
For SS 433 we fitted for the two lobes using HAWC and LHAASO data and also did another fit for the total summed up data reported by LHAASO. We did two independent fittings for the two lobes to allow for possible variations in particle spectrum or transport properties. Thus, we tried to find the hadronic contribution from those regions..

For each fit, we report the values of the maximum a posteriori (MAP) parameter and the corresponding $68\%$ (equivalent to $1\sigma$) credible intervals, shown in table \ref{tab:model_parameters}. The resulting posterior samples were then flattened and also visualized with corner plots shown in the appendix \ref{sec:corner}. The corner plots show the marginalized 1D distribution of each parameter with 2D correlations between the parameters. The median of each distribution was taken as the best-fit value. The 16th and 84th percentiles defined the $68\%$ credible intervals. We have included an exponential cutoff $\exp[-(E_\gamma / (0.1\, E'^{\max}_p)^{s}]$ with s=2.5 for the northern site of V4641 Sgr and s=1 for all other fits. Initially we took $\Gamma$ as a free parameter in the MCMC fit but the resulting posterior distributions were poorly constrained and highly degenerate with other free parameters. This means that the available data are not sensitive enough to constrain $\Gamma$, which causes the sampler to explore physically unrealistic regions. Therefore, we performed independent fits for fixed $\Gamma$ values 1-10 (plausible range for Galactic microquasar jets). Based on the given parameter values listed in table \ref{tab:model_parameters} and the prior ranges, discussed in section \ref{sed}, we found $\Gamma=2.6-4$ shows well-behaved posterior distributions with similar maximum log-likelihood and minimum $\chi^2$ values. For the total summed emission fit of SS 433 \ref{fig:ss433}(a), we found $\Gamma=1.03 - 1.4$. However, for the east and west lobes, we got $\Gamma=1.03 - 1.1$. Our results are thereby consistent with previously reported values $\Gamma_{\text{min}}^{\mathrm{V4641\ Sgr}}=2.6$ \cite{miller2006opening} and $\Gamma_{\text{min}}^{\mathrm{SS\ 433}}=1.03$ \cite{escobar2022highly}.\\

\begin{figure}[hbt!]
		\centering
		\includegraphics[scale=0.6]{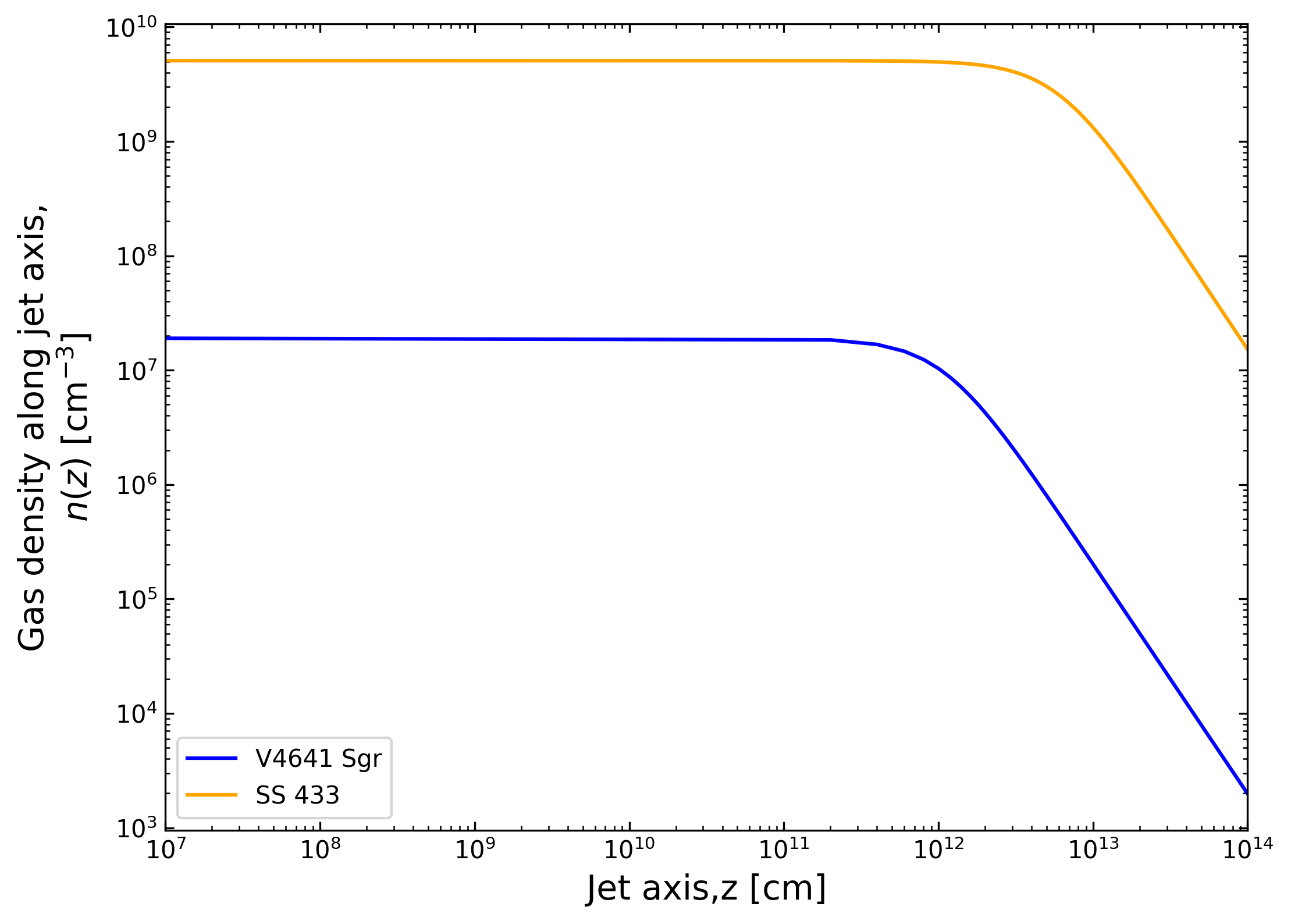}
\caption{Gas density profile along the jet axis for the microquasars V4641 Sgr (blue line) and Ss 433 (orange line). The density, $n(z)$ as discussed in Section~\ref{model} is computed for $z_0 \le z \le z_{max}$, where $z_0$ is base of the jet (values listed in table \ref{tab:model_parameters}) and $z_{max}$ is the jet extension, taken up to $10^{14}\,\cm$. Other parameters adopted to calculate $n(z)$ are listed in table~\ref{tab:model_parameters}}.
    \label{Fig:gasdensity}
\end{figure}

Using the parameter values listed in table \ref{tab:model_parameters}, we calculated the gas density flowing in the base of the jet, $n_0$, as discussed in section \ref{model} and found the values to be $1.90\times10^{7}\ \mathrm{cm^{-3}}$ and $5.08\times10^{9}\ \mathrm{cm^{-3}}$ for V4641 Sgr and SS 433 respectively. We also estimated values of the constants in injection proton spectra at $z_0$, listed in table \ref{tab:model_parameters}. Such a high gas density indicates a substantial population of relativistic protons near the jet base which can effectively contribute to the hadronic process for the production of VHE to UHE gamma-rays. Furthermore, the gas density along the jet axis, $n(z)$, determined based on the stellar wind of the companion star (see section \ref{model}) is shown in figure \cref{Fig:gasdensity}. This figure shows the gas density profile along the jet axis for both sources.

The LHAASO Galactic diffuse gamma ray data are well explained by our combined model fitting, discussed in section~\ref{diff gamma}. The best-fit values and posterior distributions of parameters of the microquasar component are shown in the appendix \ref{sec:corner}.
\section{Summary and Conclusions}
In this work, we studied the hadronic emission from two Galactic microquasars, V4641 Sgr and SS 433, which have been recently detected in multi-TeV gamma-rays by several observations. We have modeled these observed TeV gamma-ray emissions within a hadronic framework on the basis of $P-P$ interactions in the central binary and jet environment. The hadronic scenario considers the interaction of jet relativistic protons with the stellar wind. Using the Bayesian MCMC fitting of observational data from LHAASO, HAWC and HESS, we constrained the physical parameters of the microquasar jet- the hadronic jet power fraction ($q_j$), the spectral index of accelerated protons ($\alpha$), the maximum proton energy ($E'^{\max}_p$) and the jet bulk Lorentz factor ($\Gamma$). For both sources, our best-fit solutions show that the protons are accelerated in PeV energies, which implies the MQs are the efficient Galactic PeVatron candidates. The resulting hard proton spectra, high gas density and the jet-disk environment of these systems allow for a hadronic contribution from the $P-P$ interactions and produces TeV gamma-rays consistent with the observed gamma-ray emission. The predicted neutrino flux from this hadronic model reaches of the order of $10^{-12} - 10^{-14} \mathrm{erg \cm^{-2}s^{-1}}$ in TeV range, consistent with the result reported by \cite{peretti2025particle}. This essentially suggests V4641 Sgr as a promising target for KM3NeT-ARCA and TRIDENT, $5\sigma$ discovery flux of 10 years data. On the other hand, SS 433 may be detectable with longer exposures. Further, to evaluate the contribution of MQs as Galactic diffuse TeV gamma-ray emission, we generated a synthetic population of Galactic MQs with parameters sampled from the physically motivated prior ranges, consistent with our best-fit solutions of hadronic models. We compared our population modeled flux including other contributions from Molecular clouds with the observed diffuse gamma-ray emission for both the inner and outer Galaxy. Our results show that approximately 14 Galactic MQs with the jet and environmental properties similar to V4641 Sgr and SS 433 could be sufficient to reproduce the observed diffuse TeV gamma-ray emission at the inner and outer Galaxy with maximum proton energy up to 10 PeV. This suggests that a population of Galactic MQs could significantly contribute to the Galactic diffuse gamma-ray emission and act as efficient accelerators of Galactic cosmic rays up to the knee region. The maximum energies of accelerated protons for the two MQs we have discussed is between 1 and 5 PeV. The observed diffuse Galactic TeV–PeV gamma-ray data suggest that protons in microquasars can be accelerated beyond 1 PeV. Continued observations from both gamma-ray and improved neutrino telescopes would be the key to confirm the MQs as Galactic cosmic ray sources.

\bibliography{ref_prop}
\bibliographystyle{JHEP}

\section{Appendix}
\label{sec:corner}
\begin{figure}[h!]
    \begin{subfigure}[b]{0.4\textwidth}
        \centering
        \includegraphics[width=\textwidth]{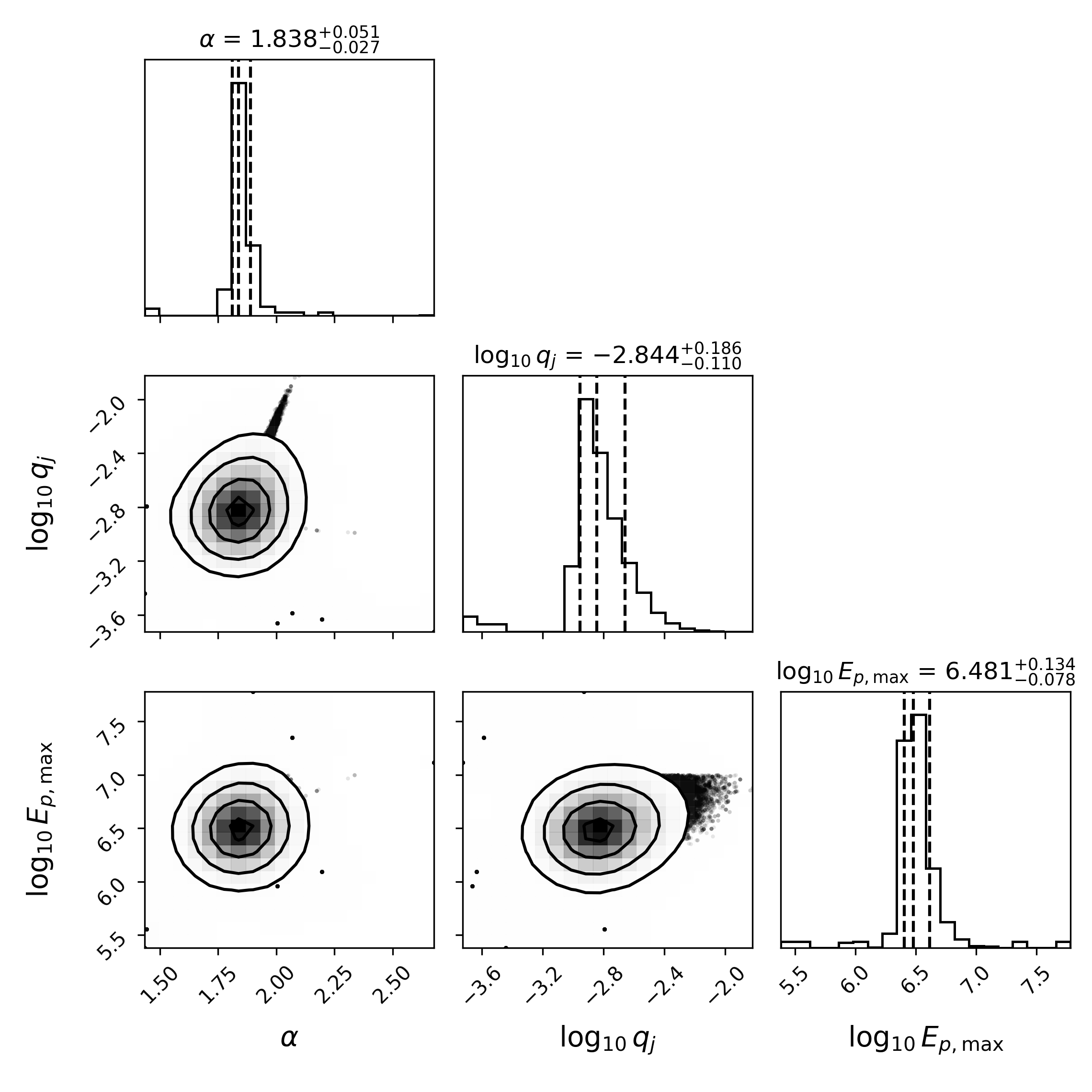}
        \caption{V4641 Sgr}
    \end{subfigure}
    \hfill
    \begin{subfigure}[b]{0.4\textwidth}
        \centering
        \includegraphics[width=\textwidth]{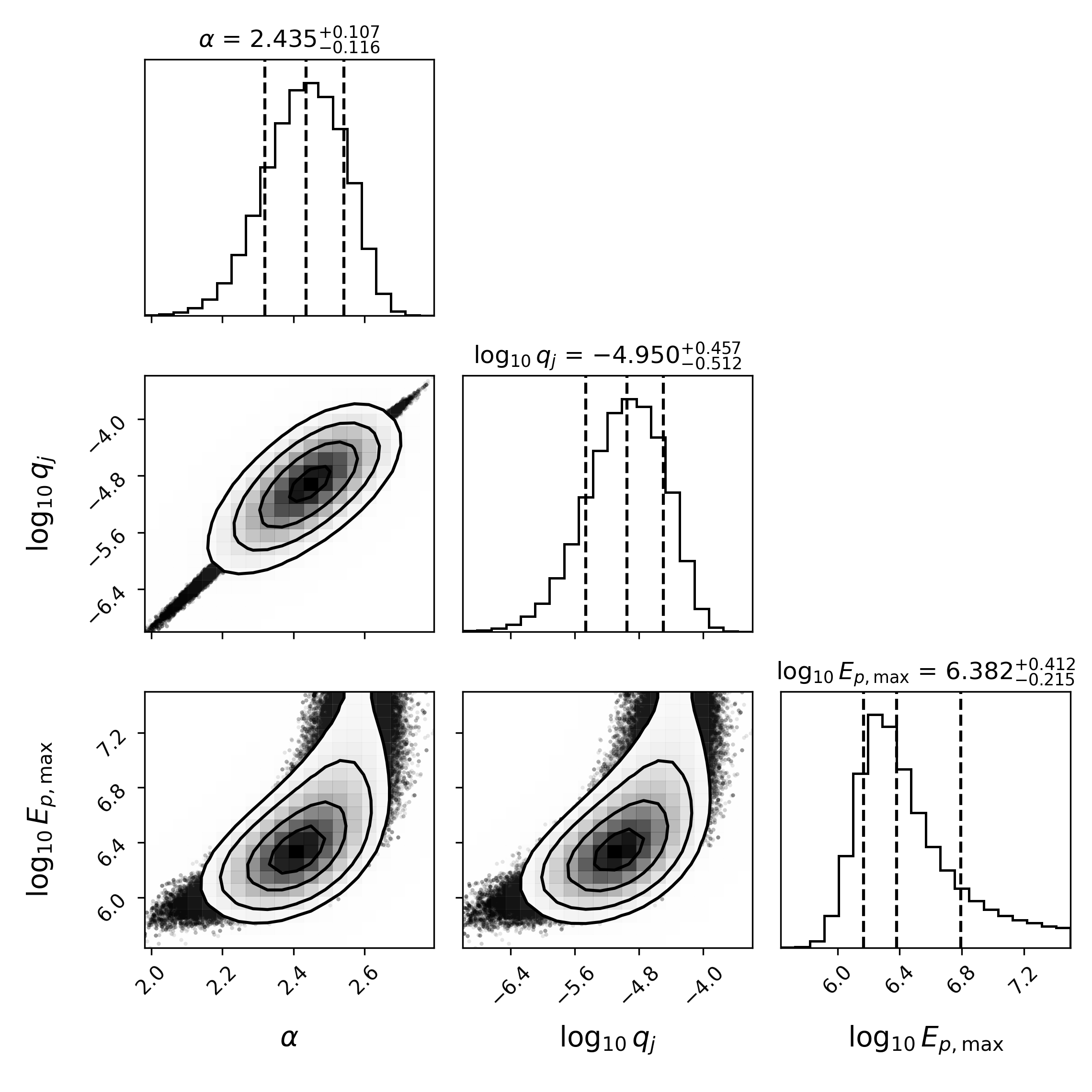}
        \caption{SS 433}
    \end{subfigure}
    \caption{Posterior distributions of free parameters of the hadronic model for the production of gamma-rays from V4641 Sgr (fig. a) and SS 433 (fig. b). The vertical dashed lines indicate median values with $68\%$ credible intervals for each distribution.}
    \label{fig:corner}
\end{figure}

\textbf{A. Posterior Distributions and Corner Plots}\\
[2pt]
In table \ref{tab:model_parameters}, we have shown our best-fit parameter values, estimated from the MCMC fits of the hadronic model to the observed gamma-ray data from MQs. Here we present the posterior distributions of free parameters of the model - the hadronic jet power fraction ($q_j$), the proton spectral index ($\alpha$) and the maximum proton energy ($E'^{\max}_p$) with the corner plots in fig \ref{fig:corner}. Fig(a) shows for the single extended source V4641 Sgr and fig(b) for the total summed emission from SS 433. The histograms show the marginal distribution and the contour plots represent the joint distributions between the free parameters. The distributions for each parameter are characterized by their medians and $68\%$ credible interval, shown by black vertical dashed lines. The parameter values and their uncertainties are indicated in each panel of the figures.

For Galactic diffuse gamma-ray fitting, we also did MCMC and here we present the posterior distributions and best-fit values of parameters for both inner and outer galaxy fittings.

\begin{figure}[h!]
    \begin{subfigure}[b]{0.48\textwidth}
        \centering
        \includegraphics[width=\textwidth]{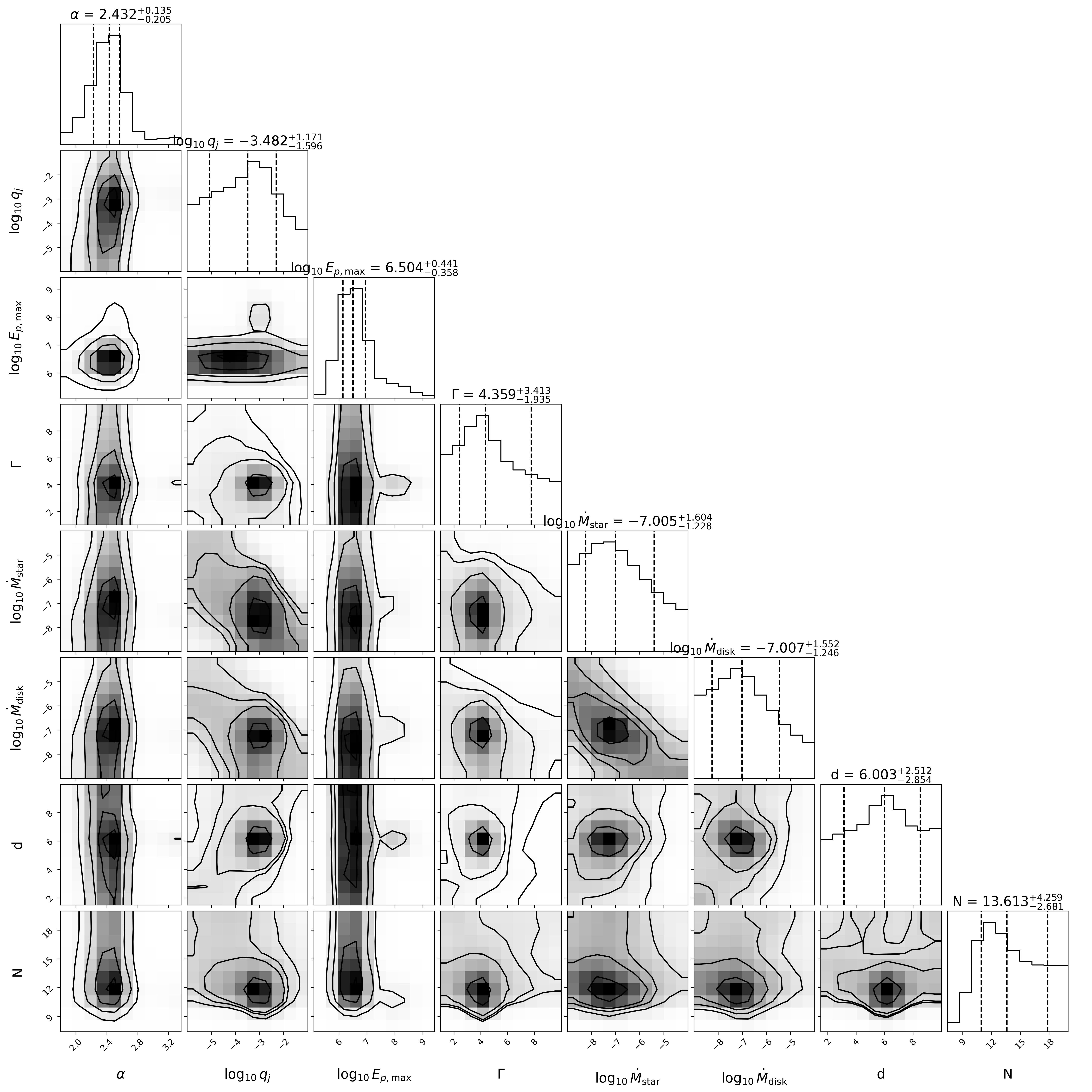}
        \caption{Inner Galaxy}
    \end{subfigure}
    \hfill
    \begin{subfigure}[b]{0.48\textwidth}
        \centering
        \includegraphics[width=\textwidth]{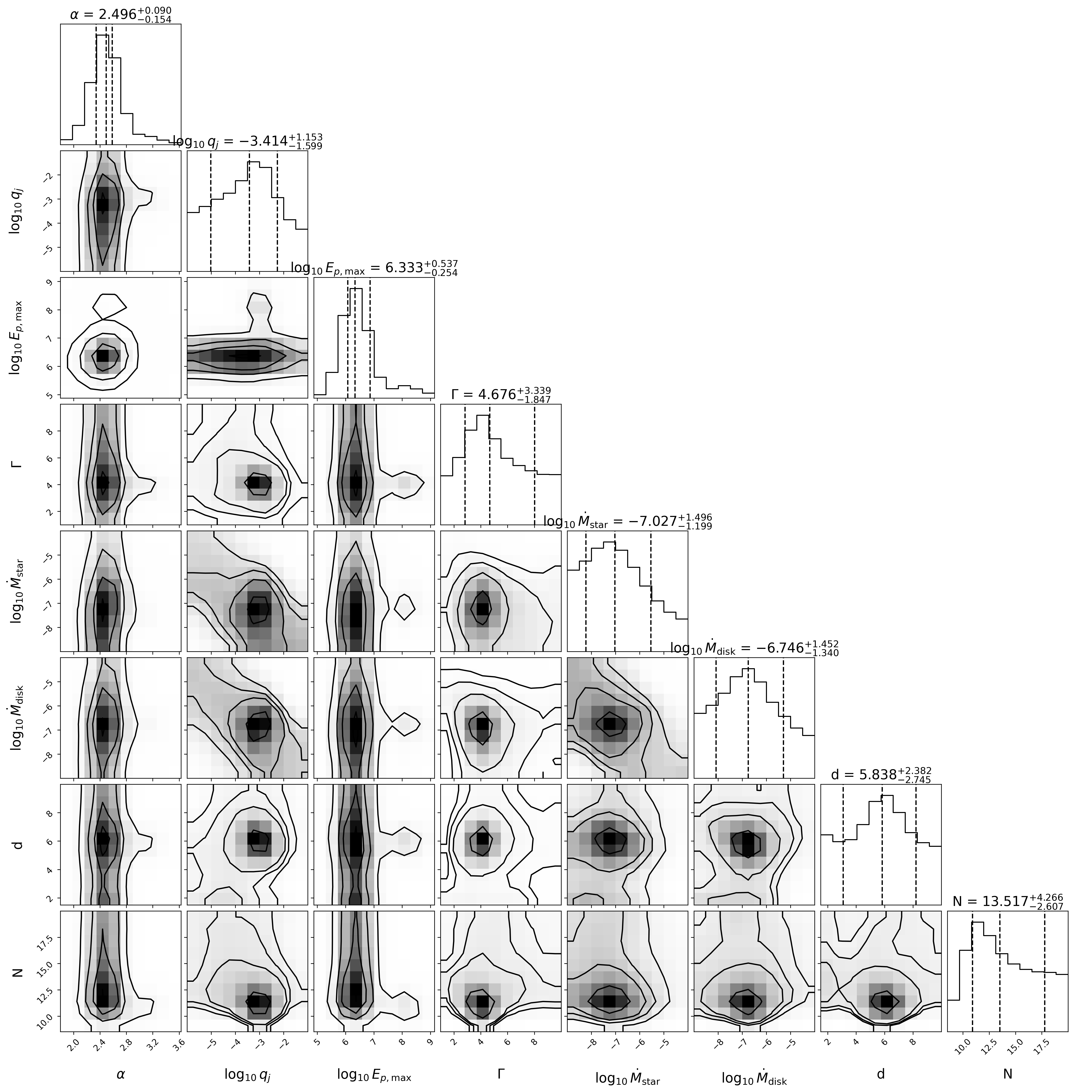}
        \caption{Outer Galaxy}
    \end{subfigure}
    \caption{Posterior distributions of free parameters of the hadronic model for the production of TeV gamma-rays from a population of Galactic MQs within the inner galaxy (fig. a) and outer galaxy (fig. b). The vertical dashed lines indicate median values with $68\%$ credible intervals for each distribution.}
    \label{fig:pop corner}
\end{figure}

\begin{table*}[t]
\centering
\resizebox{\textwidth}{!}{
\begin{tabular}{lccc}
\toprule
\textbf{Model parameters} & \textbf{Best fit values}  \\
\midrule
& \textbf{Inner Galaxy} & \textbf{Outer Galaxy} \\
Stellar mass loss rate: $\log_{10}\dot{M}_\ast$($\,M_\odot\,\mathrm{yr}^{-1}$) & $-7.01^{+1.60}_{-1.23}$ & $-7.03^{+1.50}_{-1.20}$ \\
Black hole accretion rate: $\log_{10}\dot{M}_{\mathrm{disk}}$($\,M_\odot\,\mathrm{yr}^{-1}$) & $-7.01^{+1.55}_{-1.25}$ & $-6.75^{+1.45}_{-1.34}$ \\
Distance to Earth: $d$($~\mathrm{kpc}$) & $6.00^{+2.51}_{-2.85}$ & $5.84^{+2.38}_{-2.75}$ \\
Jet bulk Lorentz factor: $\Gamma$ & $4.36^{+3.41}_{-1.94}$ & $4.68^{+3.34}_{-1.85}$ \\
Injection proton spectral index: $\alpha_p$ & $2.43^{+0.13}_{-0.21}$  & $2.50^{+0.09}_{-0.15}$ \\[2pt]
Jet-disk coupling constant: $\log_{10}q_j$ & $-3.48^{+1.17}_{-1.60}$ & $-3.41^{+1.15}_{-1.60}$ \\[2pt]
Maximum proton energy: $E'^{\max}_p$(PeV) & $3.19^{+4.38}_{-0.80}$ & $2.15^{+5.27}_{-0.95}$ \\[2pt]
No. of MQs & $13.61^{+4.26}_{-2.68}$ & $13.52^{+4.26}_{-2.61}$ \\[3pt]
\hline
\end{tabular}
}
\caption{Best-fit Hadronic gamma-ray emission model parameters for MQs to explain Galactic diffuse TeV Gamma-ray emission.}
\label{tab:popmodel_parameters}
\end{table*}

\end{document}